\documentclass[aps,prl,twocolumn,showpacs,groupedaddress]{revtex4}

\usepackage{graphicx}  
\usepackage{dcolumn}   
\usepackage{bm}        
\usepackage{amssymb}   
\usepackage{amsmath}
\usepackage{float}
\usepackage{color}

\begin{document}

\title{Atom interferometry with the Sr optical clock transition}

\author{Liang Hu}
\altaffiliation[Also]{ ICTP, Trieste, Italy.}
\author{Nicola Poli}%
\altaffiliation[Also]{ CNR-INO, Firenze, Italy.}
\author{Leonardo Salvi}%
\author{Guglielmo M. Tino}%
\altaffiliation[Also]{ CNR-IFAC, Sesto Fiorentino, Italy}
\email{Guglielmo.Tino@fi.infn.it}
\affiliation{Dipartimento di Fisica e Astronomia and LENS - Universit\`{a}
di Firenze, INFN - Sezione di Firenze, Via Sansone 1, 50019 Sesto Fiorentino, Italy}

\date{\today}

\begin{abstract}
We report on the realization of a matter-wave interferometer based on single-photon interaction  on the ultra-narrow optical clock transition of strontium atoms.  
We experimentally demonstrated its operation as a gravimeter and as a gravity gradiometer. No reduction of interferometric contrast
was observed up to an interferometer time $2T=10$ ms, limited  by geometric constraints of the apparatus. In the gradiometric configuration, the sensitivity approaches the shot noise limit. Single-photon interferometers represent a new class of high-precision sensors that could be used for the detection of gravitational waves in so far unexplored frequency ranges and to enlighten the boundary between Quantum Mechanics and General Relativity.

\end{abstract}

\maketitle


Matter-wave interference enables the investigation of physical interactions at their  fundamental quantum level and forms the basis of high-precision inertial sensors and for application in precision gravitational field sensing \cite{tino2014atom}. Today's best atom interferometers, based on multi-photon Raman/Bragg transitions and Bloch oscillations, can tailor matter-waves at will, up to macroscopic scales \cite{kovachy2015quantum}, preserving their coherence for extremely long times \cite{Poli2011}, allowing precision measurements of the Newtonian gravitational constant \cite{rosi2014precision}, Earth gravity acceleration \cite{peters1999measurement, charriere2012local, hu2013demonstration, mazzoni2015large}, gravity gradients \cite{mcguirk2002sensitive, sorrentino2014sensitivity,d2016bragg} and gravity curvature \cite{rosi2015measurement,asenbaum2017phase}. At the same time, optical spectroscopy of ultra-narrow optical transitions in atoms and ions have produced clocks with the highest relative frequency accuracy,  approaching the $10^{-19}$ level \cite{Poli2013, Ludlow2015, ushijima2015cryogenic, campbell2017fermi}. Thanks to these impressive results, schemes for gravitational waves detectors based on atom interferometers and optical clocks have been proposed
\cite{tino2007possible, dimopoulos2008atomic,yu2011gravitational, graham2013new, kolkowitz2016gravitational}. 

%
In this Letter, we demonstrate an atom interferometer based on the ultra-narrow $^1$S$_0$-$^3$P$_0$ optical clock transition of $^{88}$Sr atoms. 
While atom interferometry with the 400~Hz-wide intercombination transition of calcium  was reported already \cite{riehle1991optical, ruschewitz1998sub, sterr2004optical}, the virtually infinite lifetime of the upper clock state in strontium is crucial for demanding applications as, for example,  gravitational wave detectors. 
Based on a single-photon process,  
this novel sensor will indeed enable new fundamental tests, lying at the border of Quantum Mechanics and General Relativity \cite{margalit2015self, amelino2014gravity}, such as quantum interference of clocks, with the possible observation of the red-shift induced decoherence effects \cite{zych2011quantum,Pikovski2015}, light Dark Matter search \cite{geraci2016} and tests of the Weak Equivalence Principle with quantum superpositions of states with large energy ($\sim$eV) separation \cite{rosi2017quantum, tarallo2014test}. Precision measurements of gravity will also be necessary for the development of optical lattice clocks. The comparisons of clocks at  the $10^{-19}$ level will not only require a precise knowledge of the static gravitational field component at the atomic cloud location, but it will also demand a simultaneous measurement of time-varying gravitational potential effects down to the exceptional level of
$10^{-2}$~m$^2$/s$^2$ \cite{voigt2016time}. Novel single-photon interferometers, based on the same atomic transition employed as optical frequency reference in  the clocks, will then play an important role in this field by enabling advanced experimental sequences with interleaved precision measurements of the optical frequency and the gravitational potential.

\begin{figure}
    \centering
    \includegraphics[width=1\linewidth]{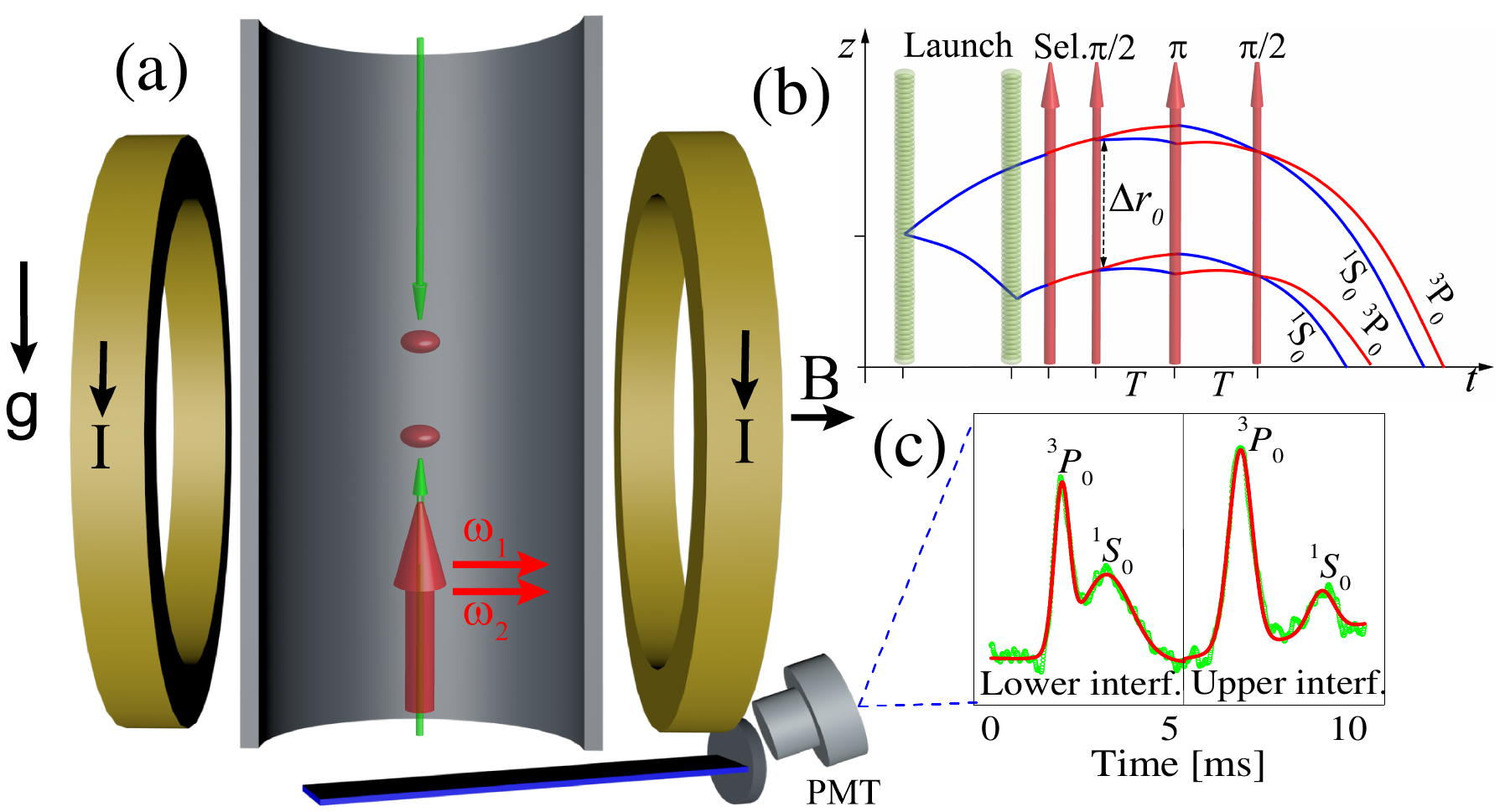}
    \caption{Schematic view of the single-photon gradiometer setup (not to scale). (a) After the MOT cooling stages, two $^{88}$Sr clouds are launched upwards in a fountain through accelerated optical lattices at 532 nm (green arrows). During the free-fall, the clouds are velocity selected by $\pi$-pulses on the clock transition followed by a blow-away laser pulse at 461 nm that eliminates the residual hot non-selected atoms. A standard Mach-Zehnder sequence of clock pulses is then applied (red arrows), where an artificial phase shift $\delta\phi$ is induced by changing the phase shift between the two  frequencies ($\omega_{1}$ and $\omega_{2}$) used to interrogate the two clouds, separated by a vertical distance $\Delta r_0$. The clock transition is induced by a static magnetic field $B$ \cite{taichenachev2006magnetic} generated by the MOT coils. The direction of $B$ is parallel to the polarization of the two laser fields.  
    After repumping the excited atoms back into the ground state, the relative population in the two arms of each interferometer is detected by collecting the fluorescence signal produced by a resonant sheet of blue light at 461~nm onto a photomultiplier tube (PMT). (b) Space-time trajectories of atoms in the single-photon gradiometer. (c) Typical fluorescence signals at the output of the single-photon gradiometer.}
    \label{fig:fig1}
\end{figure}

%
For a single-photon interferometer, the phase shift $\Delta\varPhi$ between the two interferometer arms is given by \cite{antoine2006matter,graham2013new} (retaining only phase corrections up to second order in the ratio of $\pi/2$-pulse duration $\tau$ to the interferometer pulse separation $T$):
\begin{equation}
\begin{split}
\Delta\varPhi = &\left(\alpha-g\,\omega_{a}/c\right)T^{2}\left(1+\frac{2\tau}{T}+\frac{4\tau}{\pi T}+\frac{8}{\pi}\left(\frac{\tau}{T}\right)^{2}\right)\\
&+(\phi_{1}-2\phi_{2}+\phi_{3}),\\
\end{split}
\label{eq:1}
\end{equation}
where $\alpha$ is the frequency chirping rate applied to the clock pulses, necessary to compensate for the gravity-induced Doppler shift and $\hbar\omega_{a}$ is the atomic energy difference between the two clock states. The quantities $\phi_{i}=\omega_{a} z_{i}/c-\omega t_{i}$ (where $\omega=2\pi c/\lambda$ is the clock laser frequency) are the phases of the clock laser field at positions $z_{i}$ and times $t_{i}$ at the beginning of each optical pulse. 

The practical implementation of single-photon interferometers presents several challenges.  High-power and high frequency stability laser systems are required to drive the high-$Q$ optical clock transition with sufficiently high Rabi frequencies. This condition is even more stringent in the case of multi-photon large momentum transfer (LMT) beam splitters \cite{graham2013new, parker2016controlling}. As Eq.\ref{eq:1} shows, single-photon interferometers are sensitive to the  optical phase of the clock laser field, that is, to the phase difference of the laser field with respect to the atomic phase, rather than to the phase difference between  two laser  fields, as in two-photon Raman/Bragg interferometers. This  sets stringent requirements on the  phase stability of the laser employed to drive the clock transition. In this case, for instance, a small fluctuation in the optical path  $\Delta l\sim200$~nm between the clock laser and the atoms, induces a phase change of $\Delta\phi_{i}=2\pi \Delta l/\lambda>\pi/2$, that is large enough to mask the interferometer signal. Therefore, to exploit the benefit of using the clock transition in atom interferometry, techniques to actively cancel the phase noise of the optical field or differential interferometric schemes need to be employed. 


\begin{figure}
	\centering
	\includegraphics[width=1\linewidth]{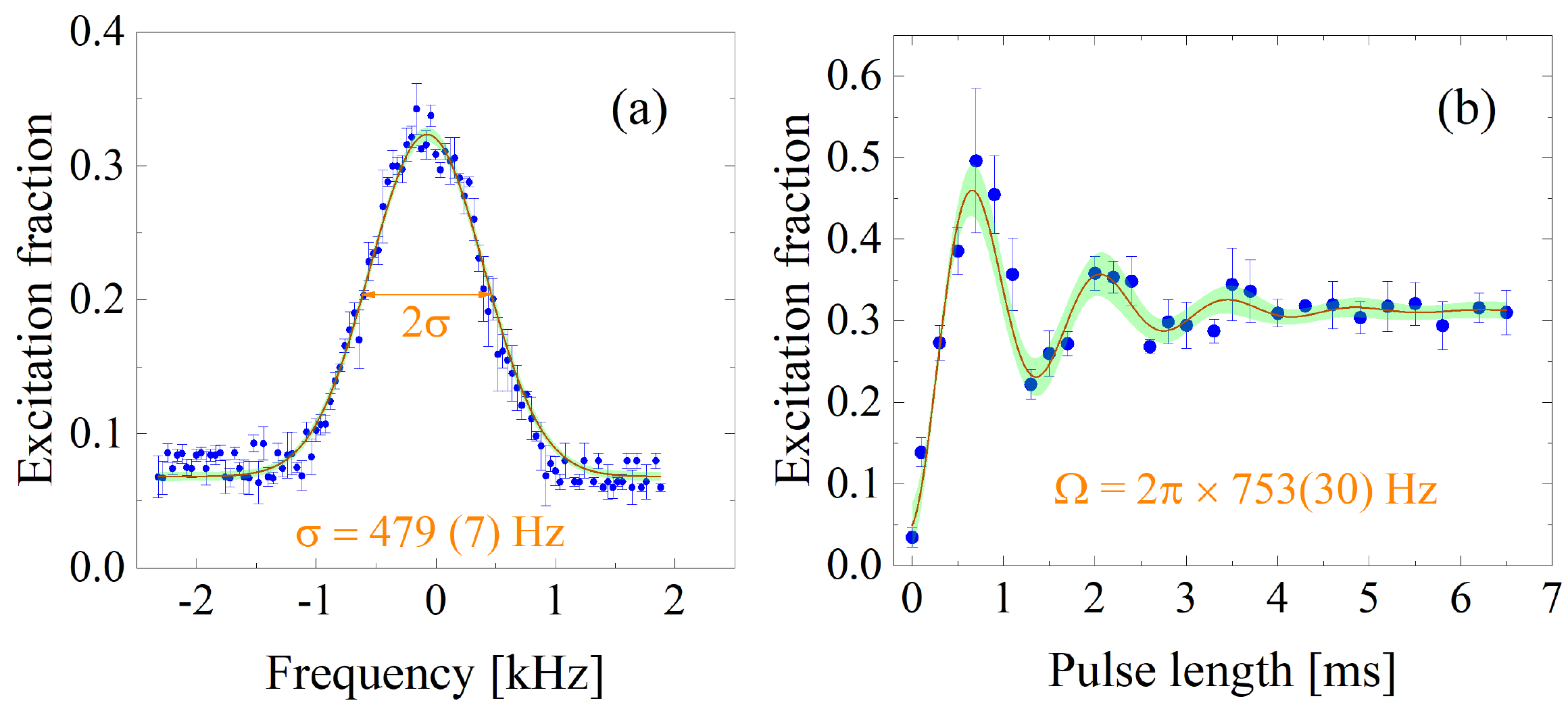}
	\caption{(a) Clock spectroscopy signal on the free-falling, momentum
			selected cloud. The signal linewidth, by eliminating the effect of the finite spectroscopy pulse length (1.2 ms) and estimated from a Gaussian fit of the profile $\sigma$, indicates an atomic momentum
			spread of $0.04~\hbar\omega_{a}/c$. (b) Rabi oscillations of 
		atomic excitation fraction as a function of clock pulse length. The oscillation is recorded with typical values of static magnetic field
		($B$=330~G) and clock light peak intensity (20~W/cm$^{2}$). The experimental result fits well
		with a damped sinusoid with a corresponding Rabi frequency of
		$\Omega= 2\pi\times753(30)$~Hz and a damping time of $1.18(0.15)$~ms} 
	\label{fig:fig4}
\end{figure}

\begin{figure*}
	\centering
	\includegraphics[width=1\linewidth]{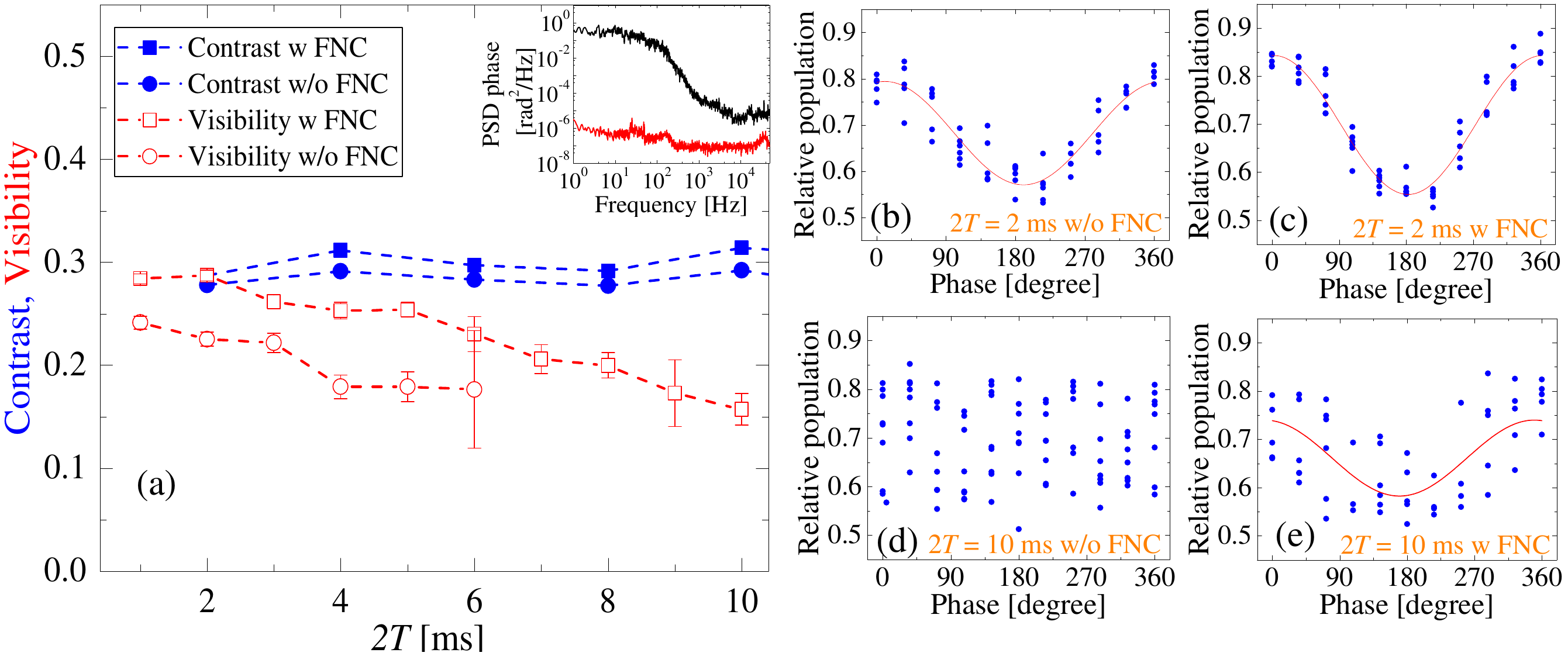}
	\caption{(a) Measured contrast and fringe visibility of a
		Mach-Zehnder single-photon interferometer (single-cloud gravimeter configuration) 
		as a function of interferometer time $2T$. Fringe visibility (red open squares and circles) is given by the	amplitude of the fitted sinusoidal function on each dataset, as shown in (b)-(e). Contrast (solid squares and circles) is estimated by data dispersion from the 2$^\mathrm{nd}$ to the 98$^\mathrm{th}$ percentile. Both contrast and visibility are measured with (squares) and without (circles) the active fiber noise cancellation (FNC) system. The inset shows the  in-loop phase noise power spectral density (PSD) of the 10~m-long optical fiber connecting the clock laser source to the atom interferometer, with (lower red curve) and without (upper black curve) FNC, respectively. Interferometer fringes in (b) and (c) are taken for $2T=2$~ms, with and without FNC respectively, while interferometer fringes in (d) and (e) correspond to $2T=10$~ms, with and without FNC, respectively.}
	\label{fig:fig2}
\end{figure*}

For the implementation of the  interferometer, we chose the most abundant bosonic isotope $^{88}$Sr, for which the forbidden $J=0-J=0$ optical clock transition can be induced and tuned in its strength by a
static magnetic field \cite{taichenachev2006magnetic}. The system used to produce ultra-cold samples of strontium has
been described elsewhere \cite{zhang2016trapped, mazzoni2015large, Poli2011}. In brief, a cloud of  $\sim 5\times10^{6}$ ultra-cold $^{88}$Sr atoms at a temperature of 1.2~$\mu$K, with horizontal (vertical) dimension of 300~$\mu$m (70 $\mu$m) at full width half maximum (FWHM), is produced by a two-stage magneto-optical trap (MOT). After the  preparation stage, about 50\% of the atoms are loaded into a vertical 1D optical lattice at 532~nm. The atoms are maintained in the trap for a time of about 65~ms, which is necessary to invert the current direction in one of the MOT coils and to turn on the homogeneous magnetic field $B$ shown in Fig. \ref{fig:fig1}(a).

The atoms are then accelerated upwards in 8~ms by frequency ramping down the top lattice beam at a rate of 100~kHz/ms, corresponding to an acceleration of $2.7~g$.
After a delay of 1~ms, the same launch procedure is repeated and the residual free-falling atoms are launched in 6~ms. Eventually, we produce two clouds with a similar number of atoms ($\sim2.5\times10^{5}$), a separation of $\Delta r_{0}=2$ mm and a velocity difference of  $\Delta v_0=24\,\hbar\omega_{a}/mc$ along the vertical axis (where $m$ is the mass of $^{88}$Sr atoms). 

To drive the optical clock transition, we employ a 1~Hz-linewidth clock laser system delivering up to 350~mW at $\lambda=698$~nm. The system consists of a master laser frequency stabilized to a high finesse cavity \cite{tarallo2011high}, and power amplified by a slave laser and a tapered amplifier. In order to simultaneously interact with both clouds, the clock laser includes two frequency components $\omega_{1}$ and $\omega_{2}$. This is implemented by feeding an acousto-optical modulator (AOM), placed on the clock beam path,  with two frequencies produced by a two-channel direct digital synthesizer (DDS) generator. The interferometer pulses at different frequencies share the same optical path including fiber, mirrors and optics, then acquiring only common-mode noise. With a 1/e$^2$ beam radius of $500~\mu$m, the typical peak intensity on the atoms is about 10~W/cm$^{2}$. 

After the launch, each cloud is velocity selected by a first $\pi$-pulse on the clock transition, which puts the selected atoms into the excited clock state $|^{3}$P$_{0},p_{0}+\hbar\omega_{a}/c\rangle$, where $p_{0}$ is the initial momentum of the atoms. Then, the atoms in the $|^{1}$S$_{0},p_{0}\rangle$ state are blown away using light resonant with the dipole-allowed $^{1}$S$_{0}$-$^{1}$P$_{1}$ transition at 461 nm. Typically, about 4\% of the atoms are selected, thus producing samples of about $10^{4}$ atoms with a narrow vertical momentum width of $\sim 0.04~\hbar\omega_{a}/c$ for each interferometer, as confirmed by the width of clock spectroscopy signal (see Fig.~\ref{fig:fig4}(a)). Indeed, this is the first demonstration of how the ultra-narrow clock transition can provide a free-falling ensemble of atoms with a very well-defined velocity, enabling high interferometer contrast \cite{szigeti2012momentum}. 


After the momentum selection, a  $\pi/2-\pi-\pi/2$ Mach-Zehnder interferometer pulse sequence based on the ultra-narrow $^1$S$_0$-$^3$P$_0$ optical clock transition is applied to both atomic clouds. In this way, atomic ``internal'' (electronic)  clock states and ``external'' (momentum) states are entangled.  At the end of the interferometer sequence, a vertical push beam at 689~nm from the bottom side   is used to decelerate atoms in the ground state for spatially separating from atoms in the excited state. After repumping the excited atoms back to the ground state, the relative population in the two arms of each interferometer is detected by collecting the fluorescence signal produced by a resonant sheet of blue light at 461~nm onto a photomultiplier tube, as shown in Fig.~\ref{fig:fig1}(c).



The main sources of phase noise in our present configuration are acoustic and sub-acoustic vibrations coupled to the atomic system and to the 10 m-long polarization maintaining  fiber used to bring the clock laser light to the atoms.  We compared the observed contrast and fringe visibility of a single-cloud Mach-Zehnder interferometer (the cloud in this case is released directly from the 1D lattice) with and without actively stabilizing the clock laser phase at the fiber end. The active fiber noise cancellation (FNC) system has a bandwidth of 50~kHz, reducing by more than 50~dB the phase noise up to 100 Hz (see the inset in Fig.~\ref{fig:fig2}(a)).  With this system, we observed a dramatic difference for the fringe visibility and contrast when the FNC is active, as shown in Fig.~\ref{fig:fig2} (a). This is particularly clear for the fringes corresponding to the longest interferometer time $2T=10$~ms (as shown in Fig.~\ref{fig:fig2}(d) and (e) compared to (a) and (b)). It is worth noting that the fringe visibility is only partially recovered by the FNC system, as shown in Fig.~\ref{fig:fig2}(e),  since some of the optical components before the atomic cloud and outside the loop introduce un-compensated phase noise.
Moreover, additional un-compensated noise comes from shot-to-shot initial momentum $p_{0}$ fluctuations of the selected clouds.

With a single-frequency clock laser peak intensity on the atomic sample of about 20~W/cm$^{2}$ and a maximum static magnetic field $B= 330$~G, the observed Rabi frequency is $\Omega= 2\pi\times753(30)$~Hz, corresponding to a $\pi$-pulse duration $\tau_\pi=0.7$~ms (see Fig. \ref{fig:fig4}(b)). While this value appears substantially larger than the typical Raman/Bragg pulses duration, it is important to notice that in the case of single-photon clock interaction, the losses due to spontaneous emission are vastly reduced. This represent a huge advantage with respect to two-photon Raman/Bragg pulses, where the spontaneous emission constitutes the main limitation to interferometer contrast and  short pulses duration and large finite detunings from the single-photon transition are typically required\cite{d2016bragg}. 

In our case, the main factor affecting the observed interferometer visibility and contrast is  the  $\pi$-pulse efficiency $\sim50$\% (see Fig.~\ref{fig:fig4}(b)) limited mainly by the initial atomic momentum width. The pulse  duration for the momentum selection is indeed a trade-off between the number of  selected atoms and the momentum width. A longer selection pulse  would lead to an increase of the $\pi$ pulse efficiency at the expense of the final atom number in the interferometer. The chosen parameters are a result of an optimization of the signal-to-noise ratio at the final detection. Further reduction of the contrast is due to clock laser intensity inhomogeneity (the clock beam size is comparable to the atomic cloud size) and to the residual motion of the atoms during the $\sim$~ms long $\pi$-pulse time. Using  the fermionic $^{87}$Sr isotope  with the same laser intensity,  the Rabi frequency would be $\Omega\sim2\pi\times5$~kHz, thus allowing a larger laser beam size or a shorter $\pi$-pulse duration.

To  demonstrate that the clock phase noise can be  rejected, a gravity gradiometer without FNC system has been implemented. In a typical experimental configuration, the differential phase-shift in the gravity gradiometer is \cite{roura2017circumventing}:
\begin{equation}
\begin{split}
\Delta\phi\approx&\frac{\omega_{a}}{c}(\Gamma T^{2})\Delta r_{0}+
\frac{\omega_{a}}{c}(\Gamma T^{2})\Delta v_{0}T+ \delta\phi\\
&\equiv\delta\phi_r+\delta\phi_v+\delta\phi,\\
\end{split}
\label{eq:eq2}
\end{equation}
where  $\Gamma$ is the gravity gradient. The first two leading phase shift terms are induced by the separation $\Delta r_0$ and the velocity difference $\Delta v_0$ between the two clouds, corresponding to $\delta\phi_r=1.4\times10^{-6}$~rad and $\delta\phi_v=5.4\times10^{-7}$~rad, respectively, that are too small to be measured with our current sensitivity.
This would result in a closed ellipse dataset when plotting the interference fringe signals of one interferometer vs. the other. Therefore, in order to analyze the data, an artificial relative phase shift $\delta\phi$ has been inserted by adding a relative phase shift directly onto the two radio-frequency signals used to drive the AOM on the clock laser beam. In this case, the observed ellipse angle will be the sum of the synthetic phase $\delta\phi$ and the other phases. Compared to other methods, in which a relative phase shift $\delta\phi$ is introduced via artificial external gradients \cite{wang2016extracting, roura2017circumventing}, here no additional fields  or changes of the pulses wavenumber are required.



\begin{figure}
    \centering
    \includegraphics[width=0.9\linewidth]{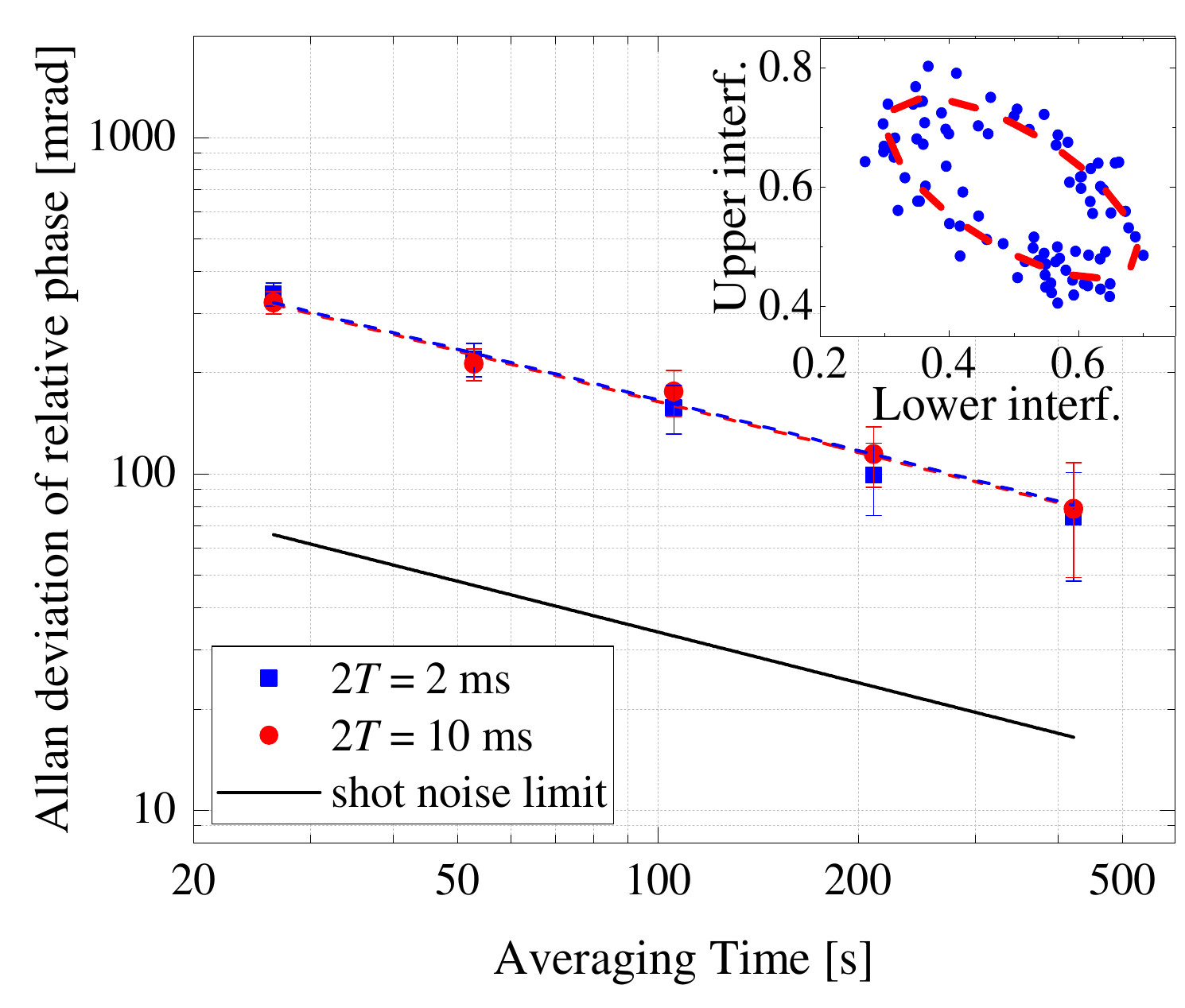}
    \caption{Allan deviation of the relative phase shift for the gradiometer
with different interferometer times of $2T = 2$~ms and $2T =
10$~ms. As expected, the gradiometer configuration rejects the
common-mode absolute laser phase noise with an Allan deviation going 
down with the square root of the averaging time, as shown by a linear fit of the data 
(dash lines). The inset indicates an
example of the relative population of the upper interferometer vs.
the lower interferometer for a time of $2T=10$~ms. From the ellipse
fitting (red dashed line) a relative phase  $\Delta\phi=2.31(0.06)$~rad can be extracted,
consistent with the value of the applied synthetic phase
$\delta\phi=3\pi/4$~rad.}
    \label{fig:ClockGradiometer}
\end{figure}

We characterized the gradiometer short-term sensitivity via 
the Allan deviation of the estimated relative phase shifts for two
independent sets of 1300~cycles (see Fig.
\ref{fig:ClockGradiometer}). The measurements have been repeated for two different 
interferometer times,  $2T=2$~ms (black
squares)  and $2T=10$~ms (blue triangles), respectively. The cycle time was set
to 2.4~s, resulting in an overall acquisition time of about 1~h, for each set. 
Each ellipse was fit on 11 successive points. For all the datasets,
as expected for cancellation of the common-mode laser phase noise, the Allan deviation scales 
as $\tau^{-1/2}$ (where $\tau$ is the
averaging time), showing that the main noise contribution comes
only from white phase noise. The relative phase sensitivity at 400~s is
70~mrad, which is five times higher than the shot noise limited
sensitivity estimated for $10^4$ atoms and with a typical contrast of $\sim30$\% (see Fig. \ref{fig:ClockGradiometer}). The current sensitivity is mainly limited by the detection efficiency at the end of the interferometer sequence.


In conclusion, we have experimentally demonstrated a novel atom interferometer based on the ultra-narrow clock transition of $^{88}$Sr atoms. Our observations illustrate the fundamental characteristics of single-photon interferometers and their  dependence on the  difference between the atomic  and the clock laser phase. We also demonstrated that it is possible to reject the clock laser phase noise by a  gradiometric configuration, thus retrieving the interferometer visibility and contrast up to interferometer times of $2T=10$ ms. The lattice double-launch technique that we implemented allowed us to add a synthetic relative phase shift $\delta\phi$ with a novel method without the need of additional external fields.


The new  atom interferometry scheme we demonstrated will be important to investigate effects at
 the boundaries of Quantum Mechanics and General Relativity \cite{zych2011quantum,rosi2017quantum}. In the future, the implementation of large momentum transfer schemes, through the use of multiple pulse configurations, and  the operation of very-long-baseline gradiometers immune  to laser phase noise might allow to detect gravitational waves in the low frequency region \cite{graham2013new,yu2011gravitational}. 
The clock transition in $^{87}$Sr is also being considered for precision atom interferometry with atoms in optical waveguides \cite{Katori2017}.

%

\begin{acknowledgments}
We thank J. Ye for a critical reading of the manuscript and R. P. Del Aguila for discussions. We acknowledge financial support from INFN and the Italian Ministry of Education, University and Research (MIUR) under the Progetto Premiale
``Interferometro Atomico" and PRIN 2015. We also acknowledge
support from the European Union's Seventh Framework Programme
(FP7/2007-2013 grant agreement 250072 - ``iSense" project and
grant agreement 607493 - ITN ``FACT" project).
\end{acknowledgments}




\end{document}